\documentclass[prb,aps,twocolumn,floatfix,psfig,epsfig]{revtex4}
\usepackage{graphicx}
\usepackage{amsmath}
\begin{document}
\title{Resonant tunneling in fractional quantum Hall effect: superperiods and braiding statistics}
\author{Jainendra K. Jain and Chuntai Shi}
\affiliation{Department of Physics, 104 Davey Laboratory, The Pennsylvania State 
University, Pennsylvania, 16802}

\begin{abstract} We study theoretically resonant tunneling 
of composite fermions through their quasi-bound states around a 
fractional quantum Hall island, and find a rich set of 
possible transitions of the island state
as a function of the magnetic field or the backgate 
voltage.  These considerations have possible relevance to a recent experimental study, 
and bring out many subtleties involved in deducing fractional braiding 
statistics.  
\end{abstract}

\maketitle

Properties of an electron system in a doubly connected 
geometry are periodic in the flux through the region devoid 
of electrons, with the period being precisely 
 one flux quantum\cite{Byers61} (flux quantum is defined as $\phi_0=hc/|e|$).
An example is Aharonov Bohm 
oscillations in the resistance of a ring as a function 
of the magnetic field.  Periods smaller than $\phi_0$ are in 
principle possible, and do occur in superconducting rings.  
It was noted in Ref.~\onlinecite{JKT}  that the fractional 
quantum Hall effect \cite{Tsui82} (FQHE) allows for the possibility 
of superperiods $K\phi_0$ with $K>1$. The geometry considered therein 
contains an island of $\nu_1$ FQHE state on the background 
of a $\nu_0$ FQHE state, and the periodic behavior occurs with respect to 
the magnetic flux through the island.  Given the singly
connected geometry, there is no reason, {\em a priori}, 
why there ought to be any period, $\phi_0$ or otherwise.  However, 
as a result of the incompressibility of the two Hall states,  
the minimal readjustment of the island is accompanied by 
a change of an integral number of flux quanta through it, 
producing, possibly, a superperiod.   
For example, for a 2/5 island surrounded by the 1/3 sea, 
a period of $5\phi_0$ was predicted\cite{JKT}.

The unusual period is closely related to the fact that 
the charge of ``quasiparticles" of the FQHE state is fractionally 
quantized \cite{Laughlin}. It is believed that FQHE quasiparticles 
also obey fractional braiding statistics \cite{Leinaas77,Wilczek82}, 
which refers to the property that the 
Berry phase associated with a closed loop of a quasiparticle
changes, upon insertion of another quasiparticle inside the loop, by 
$2\pi\theta^*$, with $\theta^*\neq$ integer \cite{Halperin84,Arovas84}.
Fractional braiding statistics was proposed as a theoretical construct  
in late 1970's \cite{Leinaas77,Wilczek82}, and 
FQHE quasiparticles are presently the only viable candidates for 
its realization. 
While interesting in its own right, a definitive 
experimental observation of 
``abelian" braiding 
statistics would also appear to be a necessary first 
step on the way to the more complicated ``non-abelian" braiding 
statistics  
believed to occur in paired composite fermion states, which has attracted much 
attention recently \cite{PT}. 

A very interesting recent experiment of Camino, Zhou, and Goldman\cite{Camino} has 
reported quasiperiodic peaks in the FQHE regime.  
They argue that the peaks occur due to 
resonant tunneling through quasi-bound states \cite{Jain89b} around 
a 2/5 island surrounded by the 1/3 sea, and estimate that the period 
corresponds to a flux change of $5\phi_0$ through the island 
as a function of the magnetic 
field, and to a change of charge two (in units of the electron charge 
$e=-|e|$) in the island as a function of the backgate potential $V_{\rm BG}$. 
They interpret this result as providing a direct observation of fractional braiding 
statistics.  The reasoning, briefly, 
is as follows\cite{Camino}.  The Berry phase 
acquired by a ``test" quasiparticle for a path encircling 
the island of area $A$ is assumed to change by $2\pi$ between two successive 
resonant tunnelings.  Modeling the test quasiparticle as an object with 
charge $e^*$ and braiding statistics $\theta^*$, this 
gives\cite{Kivelson90}, with $\phi_0^*=hc/|e^*|=(|e/e^*|)\phi_0$, 
\begin{equation}
-2\pi\frac{\Delta(BA)}{\phi_0^*}+2\pi \theta^* \Delta N_q 
= \pm2\pi\;.
\label{kivelson1}
\end{equation}
Here $\Delta(BA)$ is change in the flux and $\Delta N_q$ is 
the change in the number of quasiparticles enclosed by 
the path between two successive tunneling resonances.  
Substituting $\Delta(BA)=5\phi_0$ and
$e^*/e=1/3$ (as appropriate for a negatively charged test 
quasiparticle in the 1/3 region) gives $\theta^*=2/(3\Delta N_q)$ or 
$\theta^*=8/(3\Delta N_q)$.     
Identifying charge two with ten charge 1/5 quasiparticles in 
the 2/5 island ($\Delta N_q=10$) yields
$\theta^*=1/15$ or $4/15$, which 
was interpreted as the {\em relative} braiding statistics of 
the $e^*/e=1/3$ quasiparticle in the 1/3 state with respect to an $e^*/e=1/5$ 
quasiparticle in the 2/5 island.

Our theoretical analysis of the geometry of Ref.~\onlinecite{Camino} 
shows that,  because of a rich set of possible 
transitions of the island as well as of the island edge effects,
an interpretation of the experimental results can 
be quite subtle and complicated.  An alternative theoretical interpretation 
leads to an unacceptable value for 
the braiding statistics, and there does not seem to be any fundamental reason why 
the flux and charge cannot change by the smallest possible units.
We speculate, at the end,
on one possible scenario for reconciling theory and experiment, 
but much further work will be needed for a definitive understanding.

\underline{Transitions of a FQHE island:}
The evolution of a FQHE island is rather complicated, but 
various possibilities can be enumerated in the composite 
fermion (CF) theory\cite{Jain89}.   Composite fermions are bound states of 
electrons and an even number (taken to be two below) of 
quantized vortices, and experience a reduced effective 
magnetic field.  They form Landau-like levels in this 
reduced magnetic field, which will be called ``$\Lambda$ 
levels."\cite{lambda}  We will consider an island surrounded by the 
FQHE sea at  $\nu_0=n/(2n+1)$, which has 
$\nu_0^*=n$ filled $\Lambda$ levels.  The island state will be 
denoted by $[N_1,N_2,\cdots]$, where $N_j$ is the number of 
composite fermions in the $(n+j)^{\rm th}$ $\Lambda$ level.  In particular,
$[N_1]$ represents an island of the 
$\nu_1=(n+1)/(2n+3)$ (i.e. $\nu_1^*=n+1$) FQHE state. 
The lowest $n$ $\Lambda$ levels will be taken to be fully 
occupied and inert in the entire region of interest, and
 the number of composite 
fermions in them ($N_0$) will be suppressed for notational convenience.   
We shall assume here the simplest model,
neglecting, in particular, the effect of screening at the edge of the 
island; while one can see such  complications destroying 
simple quasi-periodic behavior, it is difficult to imagine 
how they could be responsible for it.  

\begin{table}
\caption{The change in flux, $\frac{\Delta(BA)}{\phi_0}$,
the total charge, $\Delta Q$, and the excess charge, $\Delta q$,
associated with the 
transition of the FQHE island $[N_1]$ (notation explained in text) into 
various final states.
\label{tab1}
}
\begin{tabular}{cccc}
\hline
final state & $\frac{\Delta(BA)}{\phi_0}$ & $\Delta Q$ & $\Delta q$ \\ \hline
$[N_1+1]$ & $2n+3$ & $n+1$ & $1/(2n+1)$ \\
$[N_1,1]$ & $2$ & $1$ & $1/(2n+1)$ \\
$[N_1-1]$ & $-2n-3$ & $-n-1$ & $-1/(2n+1)$  \\
$[N_1,-1]$ & $-2$ & $-1$  & $-1/(2n+1)$ \\
$[N_1-1,1]$ & $-2n-1$ & $-n$ &0 \\
$[N_1-1,n+1]$ & $-1$ & $0$  & $n/(2n+1)$ \\
$[N_1-2,2n+3]$ & $0$ & $1$ & 1 \\
\hline
\end{tabular}
\end{table}

A ``quasiparticle" of a FQHE state is an {\em isolated} composite
fermion in an excited $\Lambda$ level, and  
a ``quasihole" is a missing composite fermion from an 
otherwise full $\Lambda$ level\cite{Jain89}.  The localized charge {\em excess} 
(i.e., the charge sticking out of the uniform incompressible state; also called 
the ``local" charge\cite{Goldhaber95}) associated with a CF-quasiparticle 
of the $\nu_0=n/(2n+1)$ FQHE state is $1/(2n+1)$ in units of the electron 
charge \cite{Jain89}.   When CF-quasiparticles are far 
apart, they have well defined fractional braiding statistics 
with $\theta^*=2/(2n+1)$ (Refs. \onlinecite{Jain89,Goldhaber95}).
When they are overlapping, however, the braiding statistics 
is not a meaningful concept; that is intuitively obvious, and also 
confirmed by explicit 
calculation \cite{Kjonsberg99a,Kjonsberg99b,Jeon03}.  
The braiding statistics of CF-quasiparticles 
is not an additional concept, but a consequence of the physics encoded 
in the CF theory, and follows from the topological aspect that 
composite fermions carry an even number of quantized 
vortices \cite{Jeon03}.  The CF theory reproduces the 
earlier result \cite{Arovas84} for quasiholes at $\nu=1/3$, and enables
a microscopic evaluation of $\theta^*$  
for quasiparticles of 1/3 and other FQHE states \cite{Kjonsberg99b,Jeon03}.

What gives a discrete character to the island state is that it can change only 
through integral variations in the number of composite fermions occupying
various $\Lambda$ levels.     In deducing the corresponding
changes in the electronic state, we make use of the 
following results:  In the $\nu_1$ state: (i) 
the {\em total} charge per flux quantum is $\nu_1=(n+1)/(2n+3)$, whereas 
(ii) the {\em excess} charge per flux quantum is $\nu_1-\nu_0=[(2n+1)(2n+3)]^{-1}$.  
(iii)  From the 
perspective of the ``substrate" FQHE state ($\nu_0$), a composite 
fermion in any higher $\Lambda$ level has an {\em excess} charge of $1/(2n+1)$.

\begin{figure}
\includegraphics[width=3.0in,angle=0]{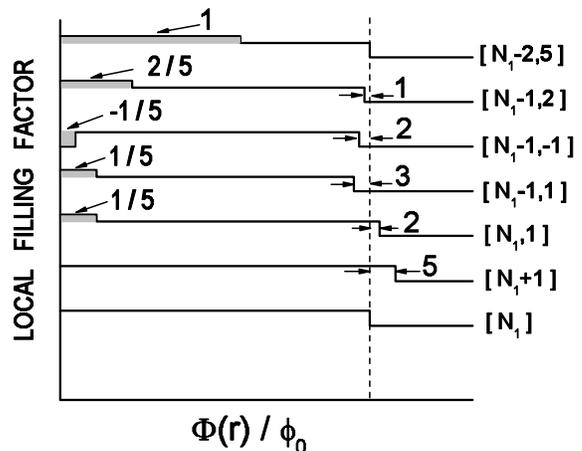}
\vspace{-8mm}
\caption{\label{fig1}  Schematic view of several transitions of the $2/5$
island discussed in the text.  The y-axis labels the local filling factor 
as a function of the distance $r$ from the center, 
with different traces offset for clarity (the filling factor on 
the far right is $1/3$ for each trace).  The x-axis is the flux through the 
area enclosed by the disk of radius $r$ ($\Phi(r)=\pi r^2 B$) 
in units of the flux quantum $\phi_0$. 
The bottom plot labeled $[N_1]$ shows a $2/5$ island inside the $1/3$ state,
with $N_1$ composite fermions in the second $\Lambda$ level.  
Addition of a composite fermion at the boundary in the second $\Lambda$ level
(at the center in the third $\Lambda$ level) 
produces $[N_1+1]$ ($[N_1,1]$), and moving a composite fermion 
from the boundary to the center results in $[N_1-1,1]$.  The top two trace
show the density for $[N_1-2,5]$ and $[N_1-1,2]$. 
The numbers near the shaded regions show 
the excess charge there.  The numbers near the 
vertical dashed line show the shift in the island edge in units of flux 
quanta.  The density oscillations at the edge and near the charge 1/5 
CF-quasiparticle have been suppressed for simplicity.  For ease of 
illustration, the island has been taken to be circularly symmetric, and all 
quasiparticles are added at the center;  changes in area and charge 
given in the text are more generally valid.
}
\end{figure}

The two ``elementary" transitions of a FQHE island are given by
 the addition of a composite 
fermion to the island edge ($[N_1]\rightarrow [N_1+1]$) or in a higher 
$\Lambda$ level in the island interior 
($[N_1]\rightarrow [N_1,1]$).  The excess charge in either case is 
$\Delta q=1/(2n+1)$, because a single CF-quasiparticle has been added.  The change 
in the total charge in the island, $\Delta Q$, and the flux passing through 
it, $\Delta(BA)/\phi_0$, are determined as follows.  In the former case
the island accommodates the excess charge by expanding to 
enclose $2n+3$ additional flux quanta, and the total charge 
increases by $\Delta Q=(2n+3)\nu_1=n+1$. 
In the latter case,  the excess charge appears in two places:  
$1/(2n+3)$ in the interior of the $\nu_1$ island, equal to the local  
charge of the quasiparticle for this state, and $2/[(2n+1)(2n+3)]$
at the boundary.  [By definition, when a composite fermion is added 
in the interior of the island, it shifts the island edge 
by an amount that encloses two additional flux quanta, 
giving an excess boundary charge of $2(\nu_1-\nu_0)$.]
The total island charge changes by $\Delta Q=1$ for the latter case,  with 
$2\nu_1$ coming from the island edge and $1/(2n+3)$ from 
the interior quasiparticle.
The $[N_1]\rightarrow [N_1,1]$ transition illustrates how apparently simple 
transitions of composite fermions manifest through rather 
complicated, nonlocal changes in the electronic state.

$\Delta(BA)$, $\Delta Q$ and $\Delta q$ can be similarly determined due to 
the removal of a composite fermion from the island edge or interior, 
or resulting from a combination of several elementary transitions.  
Many transitions are listed in Table 1 and depicted schematically 
in Fig. 1 for the special case of a 2/5 island on the 1/3 substrate.  
We have also confirmed
the basic physics presented in Fig. 1 by extensive calculations 
with explicit microscopic wave functions for composite fermions.

It may be checked that the transitions in Table 1 are, in general, 
described by the equation\cite{Kivelson90} 
\begin{equation}
-2\pi\frac{\Delta(BA)}{\phi_0^*}+2\pi \theta^* \Delta N_q 
= 2k\pi\;,
\label{kivelson2}
\end{equation}
with $\theta^*=2/(2n+1)$,  the 
braiding statistics of the CF quasiparticles of the $\nu_0$ state\cite{Jeon03},
and $\Delta N_q=r-s$ for the transition $[N_1]\rightarrow [N_1-s,r]$. 
The values of $k$ are given by $-1$, 0, 
1, 0, 1, 1, 2, for the transitions  
in Table 1, from top to bottom, respectively.  The Berry phase change between
successive resonant tunnelings can be zero (with the changes 
in the Aharonov-Bohm and the statistical phases canceling one 
another), and, under certain constraint, even $4\pi$ (an example 
is given below).

The actual island is likely to be more complicated 
than that considered above.  Depending on the shape of the 
potential due to confinement {\em and} disorder, many $\Lambda$ levels 
may be occupied and 
some localized quasiparticles 
and quasiholes may be present.  For example, 
3/7 hills or 1/3 lakes can exist inside the 2/5 island. 
The presence of such non-idealities does not affect the conclusions 
regarding changes in the size and the charge of the island due 
to addition or removal of a composite fermion at the edge or 
in the interior.  Also, transfer of a composite fermion from 
one localized state to another within the interior of the island 
does not change either its area or its charge.

\underline{Relevance to experiment:}
Returning to the experiment of Ref. \onlinecite{Camino},
the creation of a charge 1/5 particle in the 
interior of the 2/5 island has associated with it a charge 2/15 at the 
island edge, which must also be accounted for in the 
Berry phase calculation.  It is therefore more 
appropriate to view the charge two as six charge 1/5 quasiparticles in the 
island interior plus six units of charge 2/15 at the island edge.  From the vantage 
point of the test quasiparticle, this is equivalent 
to six charge 1/3 quasiparticles.  Substituting 
$e^*/e=1/3$, $\Delta(BA)=5\phi_0$, and $\Delta N_q=6$ into 
Eq. (\ref{kivelson1}) gives $\theta^*=1/9$ 
or 4/9, which now is the braiding statistics for the 
1/3 quasiparticles relative to one another.  This value 
in disagreement with the theoretically accepted one.  
Further insight is gained by noting that, for 
exterior Berry trajectories, a collection of six 
charge 1/3 quasiparticles (or, for that matter, ten charge 1/5 
quasiparticles) is topologically indistinguishable from 
two charge one electrons.  This points, on the one hand,
to a conceptual difficulty 
with interpreting the result as a measure of the relative braiding statistics 
of {\em fractionally charged} quasiparticles; on the other,   
$\Delta N_q=2$ produces $\theta^*=4/3$ 
or 1/3 for the braiding statistics of the 1/3 quasiparticle 
relative to an electron, which also contradicts 
theoretical result \cite{comment}.

From a microscopic point of view, the experimental result seems, 
at first, to correspond to the elementary transition 
$[N_1]\rightarrow [N_1+1]$ (considered in Ref. \onlinecite{JKT}) 
for which the island area increases by $5\phi_0$ and its net charge by 
two units ($\Delta Q=2$).  (Note that $\Delta N_q=1$ for this 
transition, because a single composite fermion has been added to the 
boundary of the second $\Lambda$ level.)  The value $\Delta Q=2$ appears  
to be consistent with the observed period in backgate 
voltage.  However, $\Delta V_{\rm BG}$ is proportional to the change 
in the {\em excess} charge, $\Delta q=1/3$ (table 1), and not to  
$\Delta Q$, part of which is due to the increase in the area 
of the island and counts the charge in the 1/3 substrate 
that was already present.  
A change of $\Delta q=1/3$ between two successive $\Delta V_{\rm BG}$ 
is presumably too small for the experimental parameters of 
Ref. \onlinecite{Camino}.

There, however, is no reason why the island transitions as a function 
of $V_{\rm BG}$ should be the 
same as those as a function of $B$, as assumed implicitly above.  
[Should the transitions be different, only one of 
$\Delta N_q$ and $\Delta (AB)$ in Eq. \ref{kivelson1} is known for each 
transition, preventing a determination of $\theta^*$.]  The two 
experiments ought to be analyzed independently. 
It is also necessary to allow for the possibility of more complex 
combinations of elementary transitions, which can  
occur due to energetic considerations arising
from the electrostatics of the problem.
It seems reasonable that, as a function of $B$, the charge on the 
2/5 island does not change during consecutive transitions (although 
it is redistributed internally),  because that would have  
a large Coulomb blockade energy associated with it \cite{Camino}. 
With what minimum flux change can the island
readjust while preserving the total charge inside it?
Clearly, that must involve addition of composite 
fermions in the interior of the island accompanied by removal of 
composite fermions from its edge. 
The smallest flux change is $\phi_0$, and the associated 
transition is into $[N_1-1,n+1]$ (table \ref{tab1}).   
Consider next variation of $V_{\rm BG}$.  What is the smallest unit of 
charge that can be added to the island (at a fixed $B$) 
without altering its area?  One can convince oneself that 
the transition is into $[N_1-2,2n+3]$, which 
implies $\Delta Q=\Delta q=1$.
Both these results are obvious in the 
compact spherical geometry.   The flux through a uniform density
$\nu_1$ FQHE state on a sphere (representing the island only)
can be changed by $\phi_0$ without altering the 
area or the number of particles, at the cost of creating 
quasiparticles.  Further, a unit charge can surely be added to any FQHE 
state without changing either the size or the magnetic field 
(the ``electron" goes into a 
complicated excited state of composite fermions \cite{Jain05}).    
In light of this general argument, it is not   
obvious theoretically why the flux through the island should change in 
units of $5\phi_0$ and charge in units of two.  
Incidentally, a transition $[N_1]\rightarrow [N_1-2,2n+3]$ corresponds 
to a change of Berry phase by $4\pi$ ($k=2$ in Eq.~2); this 
``superselection rule" arises from the constraint of fixed area.

We do not know, at this stage, how to reconcile theory 
and experiment.  Several parameters 
(e.g. the existence or the area of the 2/5 island;  the filling factor 
variation within the island; the relation between $V_{\rm BG}$ 
and the charge on the island) are deduced indirectly in experiment, making  
an unambiguous conversion of $\Delta B$ and $\Delta V_{\rm BG}$ 
periods into flux and charge periods difficult.  
One can therefore ask if the experiment may be consistent with 
other transitions mentioned above. It is appealing to 
consider the possibility of the smallest periods.  A $\phi_0$ 
period as a function of $B$ would imply, for the experiment of 
Ref.~\onlinecite{Camino}, an island area of 
$A=0.2 \times 10^{-8}$ cm$^{2}$, which is much smaller than either 
the lithographic area of the island ($3.5 \times 10^{-8}$ cm$^{2}$) or  
the area corresponding to the $\Delta B$ period at $\nu=1$
($1.5 \times 10^{-8}$ cm$^{2}$).  However, a smaller area could 
perhaps occur either if the 
peak density in the island is less than the density in the unpatterned 
sample, or if many disconnected patches of 2/5 exist 
(as a result of disorder) rather than a singly connected island, 
with one of them dominating 
the resonant tunneling process.  The smallest period in 
$V_{\rm BG}$ is produced by the transition $[N_1]\rightarrow 
[N_1+1]$ or $[N_1]\rightarrow 
[N_1,1]$, which slightly alters the island area but adds the smallest 
unit of excess charge ($\Delta q=1/3$).  That corresponds, for the area
$A=0.2 \times 10^{-8}$ cm$^{2}$, to   
$d\rho/dV_{\rm BG}=1.7 \times 10^8$ cm$^{-2}/V$, which compares 
favorably to the values at $B=0$ and $\nu=1$ 
($2.4\times 10^8$ cm$^{-2}/V$ and $2.0\times 10^8$ cm$^{-2}/V$,
respectively).  Obviously, further work will be needed to clarify the situation.  
It is noted that possible 
non-equilibrium or long-time scale effects have been neglected 
in our analysis.

We close with general comments regarding the relation of such a tunneling 
experiment to braiding statistics, independent of which 
transition is responsible for the quasiperiodicity.
While all transitions can be consistently 
interpreted in terms of particles with a braiding 
statistics of $\theta^*=2/(2n+1)$ (which is the braiding 
statistics of CF-quasiparticles of the substrate FQHE state), 
as evident from the discussion near Eq.~2, one may question, for 
the following reasons, if any 
of them can be taken as providing a definitive  
observation of braiding statistics.  For an unambiguous  
measurement of the braiding statistics it is crucial that the test quasiparticle be 
well separated from all of the quasiparticles it is encircling.  
For the transition $[N_1]\rightarrow [N_1+1]$, the composite fermion 
added at the edge does not qualify as a quasiparticle, but is a 
part of the $\nu_1$ state (see, for example, the $[N_1+1]$ trace in
Fig. 1).  For transitions into $[N_1,1]$ or $[N_1-1,n+1]$, the 
quasiparticles have an induced part residing at the edge, which  
interferes with the test composite fermion at the island edge, making the 
braiding statistics ill-defined\cite{Kjonsberg99a,Kjonsberg99b,Goldhaber04}. 
How about the transition $[N_1]\rightarrow [N_1-2, 2n+3]$, where the 
induced charge at the island boundary has been explicitly 
removed, leaving only $2n+3$ charge $1/(2n+3)$ 
quasiparticles in the interior?  This also does not enable 
a measurement of the braiding statistics of quasiparticles  
because, together, the interior quasiparticles behave as 
a single electron in their topological properties.  It may also be noted
that the period can be derived in all cases solely from the knowledge of 
the values of fractional charges and filling factors in the island 
and the exterior regions.

We are grateful to T.H. Hansson, S.A. Kivelson, J.M. Leinaas and S. Viefers
for numerous insightful discussions and comments. 
Thanks are due to Prof. B. Janko for his hospitality 
at the ITS, a joint institute of ANL and the University of
Notre Dame, funded through DOE and Notre Dame
Office of Research, where part of the work was completed.
Partial support of this research by the National Science 
Foundation under grant No.\ DMR-0240458 is gratefully acknowledged.

\newcommand{\journal}[1]{{#1}}
\newcommand{\PRL}{\journal{Phys.\ Rev.\ Lett.}}
\newcommand{\PRB}{\journal{Phys.\ Rev.\ B}}


\begin{thebibliography}{99}
\bibitem{lambda}
These are also called  quasi- or CF-Landau levels.
\bibitem{comment}  The relative statistics of 
a charge 1/3 quasiparticle with respect to an electron is 
$\theta^*=2$, following the arguments 
of Ref.~\onlinecite{Jeon03}, or from Eq.~2 for the bottom transition 
in Table~1.  Equating the electron to five charge 1/5 
quasiparticles allows a determination of the 
relative braiding statistics of a charge 1/3 
quasiparticle in the 1/3 state with respect to a charge 1/5 
quasiparticle in the 2/5 state to be $\theta^*=2/5$.
\end{thebibliography}
\end{document}